\documentclass[conference]{IEEEtran}
\IEEEoverridecommandlockouts
\usepackage{cite}
\usepackage{hhline}
\usepackage{amsmath,amssymb,amsfonts}
\usepackage{algorithmic}
\usepackage{graphicx}
\usepackage{textcomp}
\usepackage{xcolor}

\usepackage{multirow}
\def\BibTeX{{\rm B\kern-.05em{\sc i\kern-.025em b}\kern-.08em
    T\kern-.1667em\lower.7ex\hbox{E}\kern-.125emX}}
\begin{document}

\title{Facial Image Reconstruction from Functional Magnetic Resonance Imaging via GAN Inversion with Improved Attribute Consistency}



\author{Pei-Chun Chang\\
\IEEEauthorblockA{
\textit{Department of Computer Science} \\
\textit{National Yang Ming Chiao Tung Univ.}\\
Hsinchu, Taiwan \\
pcchang.cs05@nycu.edu.tw}
\and
\IEEEauthorblockN{Yan-Yu Tien}
\IEEEauthorblockA{
\textit{Department of Computer Science} \\
\textit{National Yang Ming Chiao Tung Univ.}\\
Hsinchu, Taiwan \\
daisy.ee08@nycu.edu.tw}
\and
\IEEEauthorblockN{Chia-Lin Chen}
\IEEEauthorblockA{
\textit{Department of Computer Science} \\
\textit{National Yang Ming Chiao Tung Univ.}\\
Hsinchu, Taiwan \\
jesy110116@gmail.com}
\and
\IEEEauthorblockN{Li-Fen Chen}
\IEEEauthorblockA{
\textit{Institute of Brain Science} \\
\textit{National Yang Ming Chiao Tung Univ.}\\
Taipei, Taiwan \\
lfchen@nycu.edu.tw}
\and
\IEEEauthorblockN{Yong-Sheng Chen}
\IEEEauthorblockA{
\textit{Department of Computer Science} \\
\textit{National Yang Ming Chiao Tung Univ.}\\
Hsinchu, Taiwan \\
yschen@nycu.edu.tw}
\and
\IEEEauthorblockN{Hui-Ling Chan}
\IEEEauthorblockA{
\textit{Center for Brain, Mind,} \\
\textit{and KANSEI Sciences Research} \\
\textit{Hiroshima Univ.}\\
Hiroshima, Japan \\
chanhl@hiroshima-u.ac.jp}
}

\maketitle

\begin{abstract}
Neuroscience studies have revealed that the brain encodes visual content and embeds information in neural activity.
Recently, deep learning techniques have facilitated attempts to address visual reconstructions by mapping brain activity to image stimuli using generative adversarial networks (GANs).
However, none of these studies have considered the semantic meaning of latent code in image space. 
Omitting semantic information could potentially limit the performance. 
In this study, we propose a new framework to reconstruct facial images from functional Magnetic Resonance Imaging (fMRI) data.
With this framework, the GAN inversion is first applied to train an image encoder to extract latent codes in image space, which are then bridged to fMRI data using linear transformation.
Following the attributes identified from fMRI data using an attribute classifier, the direction in which to manipulate attributes is decided and the attribute manipulator adjusts the latent code to improve the consistency between the seen image and the reconstructed image.
Our experimental results suggest that the proposed framework accomplishes two goals: (1) reconstructing clear facial images from fMRI data and (2) maintaining the consistency of semantic characteristics.

\end{abstract}


\section{Introduction}\label{sec:introduction}
Decoding and reconstructing what people are watching \cite{Kay2008, mozafari2020reconstructing}, imagine \cite{Reddy2010, takada2022generating}, hear \cite{chang2021decoding}, or dream \cite{Horikawa2013} from brain activity has begun receiving considerable attention recently. 
The applicability of this research area can be expanded to implementation of mind-reading technologies, which are beneficial to communicate with patients with difficult medical conditions \cite{Owen2006,Schalk2004}, or to create a sketch for artists. 
To achieve applicable social understanding and communication, accurate facial image decoding and reconstruction are especially crucial. 

Recently, with the emergence of deep learning, deep neural networks (DNNs) have shown great success in reconstructing visual stimuli from brain activity \cite{fang2020reconstructing, lin2019dcnn, st2018generative}, especially generative adversarial networks (GANs).
The GANs approach can be conceptualized as a two-player game between a generator and a discriminator.
The generator synthesizes images, and the discriminator distinguishes the real data from the fake data.
With the genearation capacity of GANs, we can easily produce a realistic image from a latent code  sampled from a specific space, such as Gaussian domain.
That is, we can reconstruct a perceived image from brain activity if we can establish the relations between brain activity and the latent code.

Shen et al. proposed an end-to-end GAN-based architecture that accepts functional Magnetic Resonance Imaging (fMRI) data as inputs and generates seen images \cite{shen2019end}.
Their method, beside comparing the stimulus image and the reconstructed image, also computed the losses of semantics in the feature domain.
Shen et al. further applied a deep generator network to optimize the reconstructed image with high-level features in the multiple layers of DNN \cite{shen2019deep}.
In particular, they assumed that visual features can be decoded from fMRI data, and that these features are similar to those calculated from the seen image \cite{horikawa2017generic}.

In fact, training a GAN model that directly maps fMRI data to images is still difficult due to the limited number of brain signal acquisitions.
Hence, VanRullen et al. assumed that there exists a transformation between image latent code
and brain signals \cite{VanRullen2019}.
The main architecture can be divided into two parts: (1) brain decoding and (2) image encoding and decoding.
Based on the variational auto-encoder (VAE), they used a discriminator for the adversarial process in the training phase to make the reconstructed image more realistic.
They first trained an image encoder or utilized principle component analysis (PCA) to transfer the image to a latent space, and applied linear regression to bridge the brain signal and the image latent code.
Finally, they decoded transferred representations to obtain reconstructed images. 
Moreover, the quality of reconstructed images probably relied on the generation ability.
They further consider using BigBiGAN \cite{donahue2019large} to replace the VAE architecture to obtain clearer images from fMRI data \cite{mozafari2020reconstructing}.

\begin{figure*}[t]
\begin{center}
	\includegraphics[clip, trim= 0cm 10cm 1cm 3cm, width=0.85\textwidth]{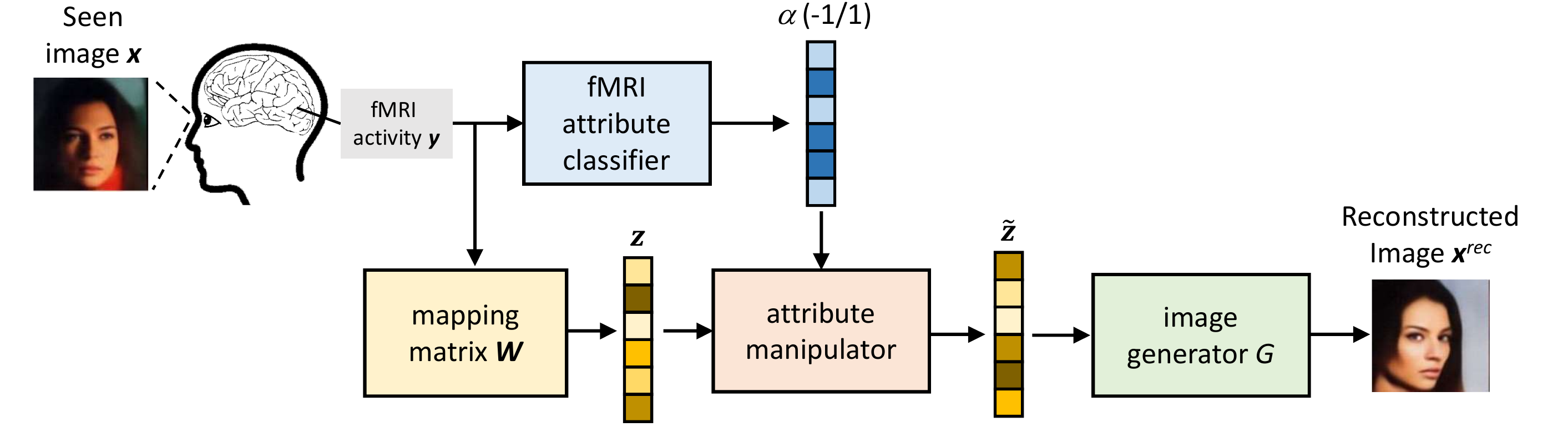}
\end{center}
   \caption{The proposed system for facial image reconstruction.}
\label{fig:overall}
\end{figure*}

However, these studies focused on how to transfer brain activity to images, and they did not consider how image latent works.
That is, there is no constraint for image space to recognize semantics (also called attributes) of the image latent.
To solve this problem, we propose a new framework for facial image reconstruction from fMRI data via GAN inversion with improved attribute consistency.
First, we apply the GAN inversion to train an encoder to extract latent codes in image space.
Then, we follow the idea from VanRullen et al. \cite{VanRullen2019} to train a simple linear transformation to convert fMRI data into image latent space.
fMRI data are also fed into several brain classifiers to determine their attributes for manipulation.
The attribute manipulator is then used to adjust the latent code to imbue it with some semantics in image space.
Finally, the styleGAN-v2 generator is used to produce the reconstructed image using the edited latent codes.

We summarize our contributions as follows:
\begin{itemize}
\item We propose a new framework to reconstruct facial images from fMRI data using GAN inversion.
\item This work is the first attempt to involve constraints for facial attributes in latent space, which gives the reconstructed image higher attribute consistency with the seen image.
\end{itemize}



\section{Materials and Methods}\label{sec:methods}

\subsection{Visual Stimulation Experiment}
A male subject, S1, participated in fMRI recordings (3T Philips ACHIEVA scanner, gradient echo pulse sequence, TR = 2s, TE = 10 ms, slice thickness = 3 mm with 0.2 mm gap, in-plane voxel dimensions $3 \times 3$ mm) during a one-back comparison task conducted in the previous study of VanRullen et al. \cite{VanRullen2019}. A total of 8440 and 20 colored face images were randomly selected from CelebFaces Attributes (CelebA) Dataset \cite{liu2015faceattributes} as the visual stimuli for training and testing data, respectively. CelebA comprises 29,264 $128 \times 128$ facial images, each of which is annotated with 40 binary attributes. In this study, three attributes with balanced ratios were selected: (1) male, (2) mouth slightly open, and (3) wearing lipstick (Table \ref{tab:img_clsfer}). 

In the one-back comparison task, each face stimulus was presented for 1 s, followed by a 2-s inter-stimulus interval. The participant pressed the button as fast as possible when a facial image was identical to the previous one. Each run consisted of 88 face trials and 30 null fixation trials, with 6-second pauses at beginning and end of each run. Among the 88 face trials, 10 face stimuli (five males and five females) were randomly drawn from the 20 testing face images and 8 trials were one-back trials, which repeatedly showed facial stimuli of the preceding trial. Note that each training face image was shown once, but each of testing face images was shown 39 to 53 times. Brain responses of one-back trials were discarded in the brain decoder training step. In the null fixation trial, a cross was presented at the center of the screen. The participant attended a total of 8 scan sessions, each of which consisted of 10 to 14 runs. 

\begin{table}
    \centering
    \caption[Ratio and accuracy for each image attribute]{Ratio of image attribute and accuracy of image attribute classifier}
    \begin{tabular}{|l|c|c|c|c|}
    \hline
    \multirow{2}{*}{\textbf{Attribute}} & \multicolumn{2}{c|}{\textbf{Training data}} & \multicolumn{2}{c|}{\textbf{Testing data}} \\ 
    \cline{2-5}
              & {\textbf{\textit{Ratio (\%)}}} & {\textbf{\textit{Acc (\%)}}} & {\textbf{\textit{Ratio (\%)}}} & {\textbf{\textit{Acc (\%)}}} \\ \hline
    Male              &  49.34 & 99.98 & 50.00 & 95.00   \\ \hline  %
    Wearing lipstick  & 41.14 & 99.55 & 25.00 & 80.00   \\ \hline 
    Mouth  slightly open  & 47.79  & 99.49 & 45.00 & 85.00  \\ \hline %
    \end{tabular}
    \label{tab:img_clsfer}
\end{table}

\subsection{Data Preprocessing}
fMRI data were first preprocessed using DPABI \cite{yan2016dpabi}, a toolbox developed on SPM12 \cite{Friston2007SPM}, in the following steps: slice timing correction, realignment, co-registration between functional and structural data, segmentation, and normalization using DARTEL \cite{Ashburner2007dartel} to non-linearly register the subject space to the MNI space with $61\times73\times61$ dimensions. Then the brain response of each trial was computed from preprocessed data using the general linear model of SPM12 \cite{Friston2007SPM} with timings of fixation, training face, testing face and one-back as regressors. For each trial, brain responses from the 4,720 voxels located in the bilateral calcarine sulcus, inferior occipital, fusiform gyrus, postcentral gyrus, and superior parietal lobule were selected and flattened to a one-dimensional array for further analysis. Parcellation of brain areas was based on automated anatomical atlas 3 (AAL3) \cite{Rolls2020aal}.


\subsection{Facial Image Reconstruction}
In this study, we sought to develop a generative model that can produce a facial image from fMRI data.
As shown in Fig.~\ref{fig:overall}, the main architecture consists of four models, including a feature mapping matrix $\emph{\textbf{W}}$, an fMRI attribution classifier, an attribute manipulator and an image generator $\emph{G}$.
First, fMRI data are used to form a latent vector $\emph{\textbf{z}}$ which is calculated by a mapping matrix $\emph{\textbf{W}}$.
Then, the attribute manipulator is used to adjust details to get the edited latent $\tilde{\emph{\textbf{z}}}$ for each selected facial characteristic using information from fMRI attribute classifiers.
Finally, the edited latent is used to generate a facial image having attribute consistency with the seen image by image generator $\emph{G}$.

\subsubsection{Facial image encoder}\label{sec:FIE}
In order to explore the relationship between brain activity and the corresponding image, we consider this problem as a GAN inversion task that seeks to find a latent code to recover the input image.
In this study, we follow the in-domain GAN approach proposed by Zhu et al. ~\cite{zhu2020domain} to train a facial image encoder to obtain the proper image latent codes for calculation of a mapping matrix $\emph{\textbf{W}}$.

\begin{figure}[t]
\begin{center}
	\includegraphics[clip, trim= 0.5cm 14.5cm 10cm 2.5cm, width=0.5\textwidth]{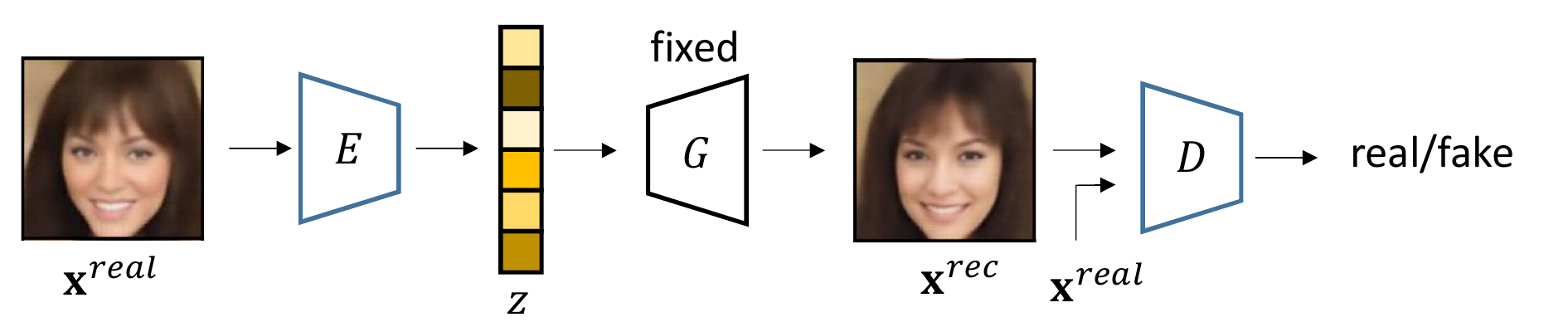}
\end{center}
   \caption{The illustration for training image latent.}
\label{fig:encoder_training}
\end{figure}

As noted by \cite{shen2020interpreting, yang2021semantic}, a well-trained GAN model can be used to encode interpretable semantics in latent space.
As shown in Fig.~\ref{fig:encoder_training}, to obtain a proper latent code, we first train an image generator $G(\cdot)$ followed by styleGAN-v2 architecture \cite{karras2020analyzing} to produce a realistic facial image.
Then, given a fixed GAN generator, the GAN inversion process attempts to discover the most accurate image latent code to reconstruct the original input image.
Hence, we fix the parameters of the generator, and train a facial image encoder to obtain the latent code in image latent space.
Encoder $E(\cdot)$ and discriminator $D(\cdot)$ are trained via minimizing loss functions $\mathcal{L}_{E}$ and $\mathcal{L}_{D}$:
\begin{equation}
\begin{split}
\min_{\Theta_{E}}\mathcal{L}_{E} = &\left \| \textbf{x}^{real} - G(E(\textbf{x}^{real})) \right \|_{2}\\
        &+ \lambda_{vgg}\left \| F(\textbf{x}^{real}) - F(G(E(\textbf{x}^{real}))) \right \|_{2}\\
        &-\lambda_{adv}E_{\textbf{x}^{real} \sim P_{data}}[D(G(E(\textbf{x}^{real})))],
\end{split}
\end{equation}
\begin{equation}
\begin{split}
\min_{\Theta_{D}}\mathcal{L}_{D} = & E_{\textbf{x}^{real} \sim P_{data}}[D(G(E(\textbf{x}^{real})))]\\
                                   &- E_{\textbf{x}^{real} \sim P_{data}}[D(\textbf{x}^{real})]\\
        &+ \frac{\gamma}{2}E_{\textbf{x}^{real} \sim P_{data}}[\left \| \bigtriangledown_{x} D(\textbf{x}^{real}) \right \|^{2}_{2}],
\end{split}
\end{equation}
where $\Theta_{E}$ and $\Theta_{D}$ are parameters of encoder and discriminator, $F(\cdot)$ is VGG feature extraction model \cite{simonyan2014very}, $\lambda_{vgg}$ and $\lambda_{adv}$ are perceptual and discriminator loss weights, $P_{data}$ is the distribution of real image data, and $\gamma$ is the hyperparameter for gradient regularization. 
In this study, the trained facial image encoder $E(\cdot)$ converts each of 128 $\times$ 128 facial images to the corresponding image latent code with 512-dimensionality.

\subsubsection{Brain Decoder}
The goal of the brain decoder is to convert fMRI data to image latent code for further facial image reconstruction.
As mentioned above, we can discover efficient latent code to reconstruct the corresponding input image using the facial image encoder.
Following \cite{VanRullen2019}, a simple brain decoder is trained to associate the facial image latent code with corresponding fMRI data.
We assume that there is a mapping matrix \emph{\textbf{W}}, which can conduct linear transformation between facial image latent codes and brain signals:
\begin{equation}
\emph{\textbf{Y}}_{N \times 4720} = \emph{\textbf{Z}}_{N \times 513} \cdot \emph{\textbf{W}}_{513 \times 4720},
\end{equation}
where \emph{\textbf{Z}} is facial image latent codes with bias term, \emph{\textbf{Y}} denotes flattened brain activity, and $N$ is the number of training facial images.
In this mathematical form, we apply pseudo inverse to estimate the optimal mapping matrix $\emph{\textbf{W}}^\mathrm{*}$:
\begin{equation}
\emph{\textbf{Z}}^\mathrm{T}\emph{\textbf{Y}} = \emph{\textbf{Z}}^\mathrm{T}\emph{\textbf{Z}} \cdot \emph{\textbf{W}},
\end{equation}
\begin{equation}
\emph{\textbf{W}}^\mathrm{*} = (\emph{\textbf{Z}}^\mathrm{T}\emph{\textbf{Z}})^{-1} \cdot \emph{\textbf{Z}}^\mathrm{T}\emph{\textbf{Y}}.
\end{equation}
With the optimal mapping matrix $\emph{\textbf{W}}^{\mathrm{*}}$, we can convert brain activity into a latent code in image space:
\begin{equation}
\emph{\textbf{Y}}\emph{\textbf{W}}^{\mathrm{*T}} = \emph{\textbf{Z}} \cdot \emph{\textbf{W}}^{\mathrm{*}}\emph{\textbf{W}}^{\mathrm{*T}},
\end{equation}
\begin{equation}
\emph{\textbf{Z}} = \emph{\textbf{Y}}\emph{\textbf{W}}^{\mathrm{*T}} \cdot (\emph{\textbf{W}}^{\mathrm{*}}\emph{\textbf{W}}^{\mathrm{*T}})^{-1}.
\end{equation}
During the testing phase, the mapped latent codes are normalized to align the mean and standard deviation calculated from the training latent \emph{\textbf{Z}}, and to discard the bias term to obtain the 512-dimension image latent code.

\subsubsection{Attribute Manipulation}
As mentioned above, we can convert fMRI data to a facial image using the mapping matrix $\emph{\textbf{W}}$ and a face image generator, but semantics (also called attributes) are not considered in these two transformations.
Following the idea of InterFaceGAN \cite{shen2020interpreting}, we apply the attribute manipulation mechanism to give the reconstructed image higher attribute consistency with the original stimulus image.

\begin{figure}[t]
\begin{center}
	\includegraphics[clip, trim= 0cm 12cm 1cm 1cm, width=0.7\textwidth]{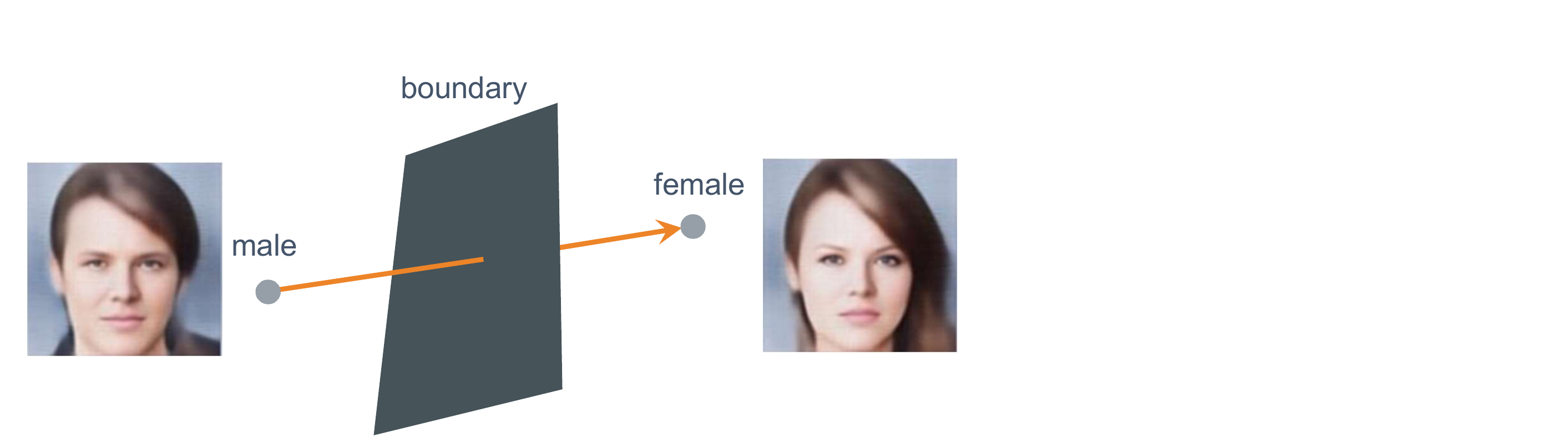}
\end{center}
   \caption{The illustration for different image latent codes across the hyperplane for gender.}
\label{fig:semantic_moving}
\end{figure}

Given a well-trained GAN model, the generator can be formulated as a function $g: \mathcal{Z} \rightarrow \mathcal{X}$.
Here, $\mathcal{Z} \subseteq \mathbb{R}^{d}$ denotes the $d$-dimension latent space, and $\mathcal{X}$ stands for the image space.
We suppose that there is a semantic scoring function $f_{S}:\mathcal{X} \rightarrow \mathcal{S}$, where $\mathcal{S} \subseteq \mathbb{R}^{m}$ represents the semantic space with $m$ attribute.
We can bridge the latent space $\mathcal{Z}$ and the semantic space $\mathcal{S}$ with $\textbf{s} = f_{S}(g(\textbf{z}))$, where \textbf{s} and \textbf{z} are the semantic scores and the sampled latent code, respectively.
According to the assumption by InterFaceGAN \cite{shen2020interpreting}, there exists a hyperplane that can be used to separate binary semantics, such as male $v.s.$ female.
In other words, as shown in Fig.~\ref{fig:semantic_moving}, semantics remain the same when the latent code walks on the same side of the hyperplane, but becomes opposite when it crosses the boundary.

Therefore, we first train the image classifiers with ResNet-34 architecture for each attribute as the semantic scoring functions.
With the pair of image latents from the facial image encoder mentioned in Sec.~\ref{sec:FIE} and the semantic score from the image classifiers, we follow \cite{shen2020interpreting} to use a support vector machine (SVM) \cite{Chang2011libsvm} to find each boundary (or called semantic direction) in image latent space to separate the attributes.
Here, for SVM training, inputs are the 512-dimension image latent codes, and labels are assigned by image classifiers.
Then, manipulative processing can be formulated as:
\begin{equation}
\emph{\textbf{x}}^{rec} = G(\emph{\textbf{z}} + \alpha \emph{\textbf{n}}) \label{editing_z},
\end{equation}
where \emph{\textbf{n}} denotes the normal direction corresponding to a specific attribute in latent space, and $\alpha$ represents the direction for manipulation.
Note that in this study, the value of $\alpha$ is $\pm1$, which represents the positive or negative direction for adjusting attributes in image latent space.

In order to determine the direction of attribute manipulation, we further train the fMRI binary classifier for each attribute.
For each attribute classifier, the input is the flatted fMRI data with 4720-dimensionality, while the output is the positive or negative attribute prediction.
As shown in Table~\ref{tab:fMRI_cls}, the architecture of the attribute classifier consists of one convolution layer, four fully-connected (FC) layers, and one linear classifier.
Each layer is followed by batch-normalization and a ReLU activation function.
To avoid training over-fitting, we apply the dropout operation with 0.5 probability for each FC-layer.
\begin{table}
\centering
\caption{The architecture of fMRI binary classifier.}
\begin{tabular}{|l|c|c|}
\hline
Name & Kernel size & Output dimensions \\ \hline
Input & & 1 $\times$ 1 $\times$ 4720    \\ \hline
Conv\_2d & \multirow{3}{*}{1 $\times$1, 8} & \multirow{3}{*}{8 $\times$ 1 $\times$ 4720}    \\
BatchNorm & & \\ 
ReLU & & \\ \hline
Flatten & & 37760 \\ \hline
FC\_1 & \multirow{4}{*}{37760 $\times$ 4720} & \multirow{4}{*}{4720}    \\
BatchNorm & & \\ 
ReLU & & \\ 
Dropout ($p=0.5$) & & \\ \hline
FC\_2 & \multirow{4}{*}{4720 $\times$ 1024} & \multirow{4}{*}{1024}    \\
BatchNorm & & \\ 
ReLU & & \\ 
Dropout ($p=0.5$) & & \\ \hline
FC\_3 & \multirow{4}{*}{1024 $\times$ 256} & \multirow{4}{*}{256}    \\
BatchNorm & & \\ 
ReLU & & \\ 
Dropout ($p=0.5$) & & \\ \hline
FC\_4 & \multirow{4}{*}{256 $\times$ 16} & \multirow{4}{*}{16}    \\
BatchNorm & & \\ 
ReLU & & \\ 
Dropout ($p=0.5$) & & \\ \hline
Classifier & 16 $\times$ 2 & 2 \\ \hline
\end{tabular}
\label{tab:fMRI_cls}
\end{table}

\subsubsection{Inference method}
As shown in Fig.~\ref{fig:overall}, to reconstruct a facial image from fMRI data, we train four models, including a feature mapping matrix, an fMRI attribution classifier, an attribute manipulator and an image generator.
First, through a standard preprocessing step, fMRI data are used to obtain the compact latent code using a feature mapping matrix \emph{\textbf{W}}.
fMRI data are also fed into the fMRI attribution classifiers to obtain the manipulation direction for each facial attribute.
Then, in the attribute manipulation step, we follow \eqref{editing_z} to adjust the latent code in image latent space, and the edited latent code is fed into the image generator to produce the corresponding facial image.

\begin{figure}[t]
\begin{center}
	\includegraphics[clip, trim= 0cm 5cm 5cm 2.5cm, width=0.7\textwidth]{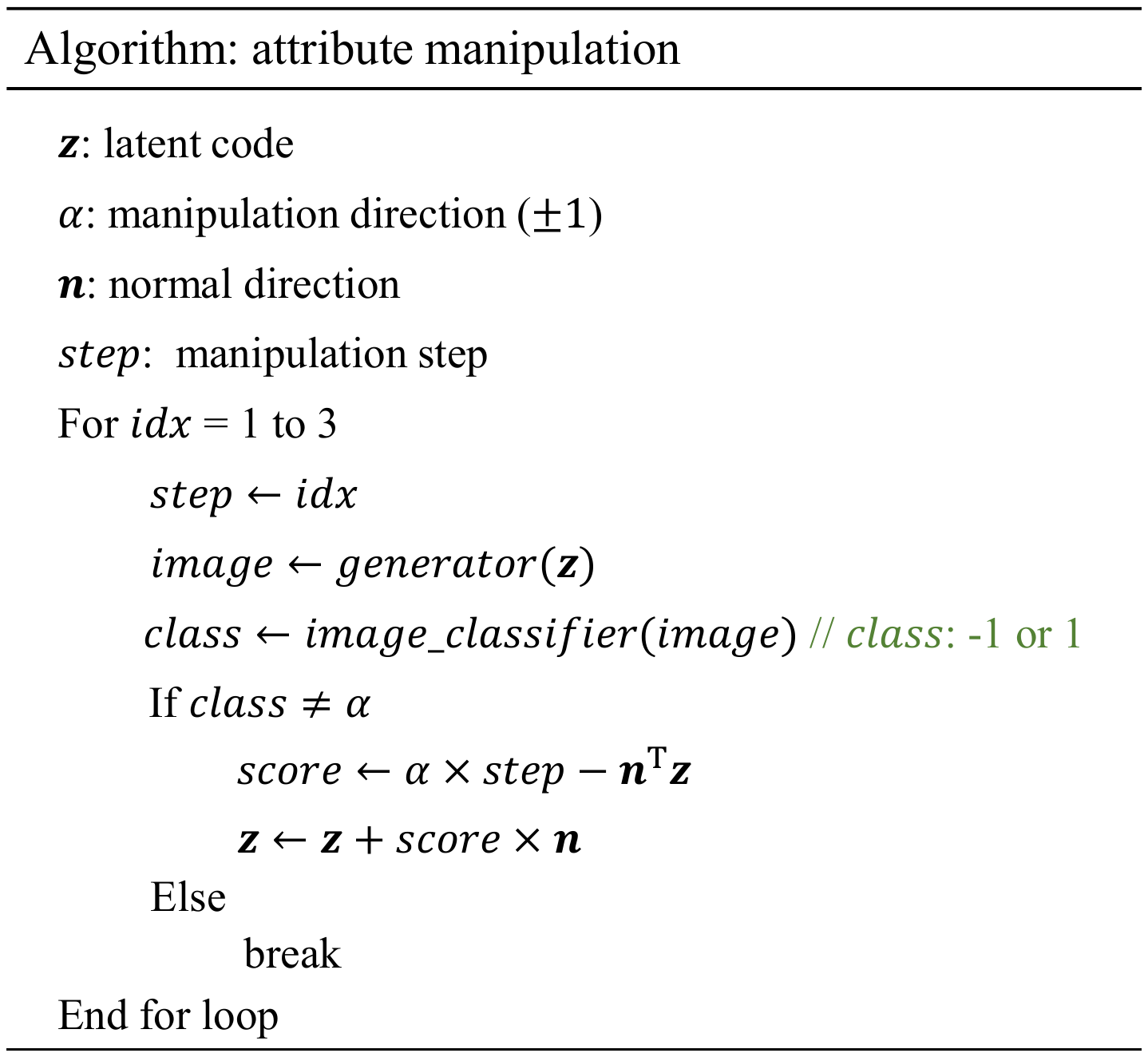}
\end{center}
   \caption{The algorithm for attribute manipulation.}
\label{fig:algorith_for_manipulation}
\end{figure}

In more detail for attribute manipulation, the direction $\alpha$ is determined by voting the results across trials with the same stimulus image.
As shown in the algorithm for attribute manipulation Fig.~\ref{fig:algorith_for_manipulation}, we first feed the original latent code into the image generator to obtain the reconstructed image.
Then, the image classifier is used to determine the facial attributes of the reconstructed image.
We further check whether the attributes of image and the direction $\alpha$ are consistent.
If the attributes of image conflict with the direction $\alpha$, we conduct the attribute manipulation with $step = 1, 2, 3$, sequentially.

\subsection{Evaluation of Attribute Classifiers}
The performance of attribute classifiers was evaluated using classification accuracy (Acc), which is the number of accurate predictions divided by the total number of predictions. Additionally, for fMRI-based attribute classifiers, the vote accuracy (Vote Acc) in classifying the testing set was computed from the number of accurate vote results divided by the total number of testing facial images. The fMRI-based attribute classifier was applied to predict the attribute for each trial of fMRI data corresponding to the targeted testing image.

\subsection{Evaluation of Reconstructed Images}
The performance evaluation of the proposed method for the reconstruction of images from fMRI data was quantified in three ways. First, to examine the quality of reconstructed images in regard to attributes, the image-based attribute classifier trained from the complete CelebA dataset was applied to classify the reconstructed images. Second, visual inspection was used to compare images reconstructed using the proposed method and those reconstructed using VAE-GAN \cite{VanRullen2019}. 
Third, we built a questionnaire for humans to judge the reconstructed images with respect to attributes and also to examine the similarity between original stimuli and reconstructed images. The questionnaire consisted of two types of questions: (1) 60 questions about attribute decisions for a reconstructed facial image and (2) 20 questions about two alternative forced choices (2AFC) between the facial image reconstructed from fMRI data corresponding to the target image and the one reconstructed from fMRI data corresponding to a non-target image. To evaluate attribute consistency between reconstructed images and original stimuli, the average accuracy (Average Acc), which is the average attribute decision accuracy of each participant, and the vote accuracy (Vote Acc), which is the average accuracy of voting results from all participants, were computed based on the feedback of attribute decision questions. The accuracy of feedback in 2AFC questions was computed to evaluate the similarity between the reconstructed and original images. Finally, feedback from 42 questionnaires was collected and the variance across all feedback was computed to check the consistency.

\section{Experimental Results}\label{sec:results}
\subsection{Performance of Facial Image Encoder and Generator}

\begin{figure}
    \centering
    \includegraphics[width=0.48\textwidth,trim=0cm 10.7cm 0 0.3cm, clip]{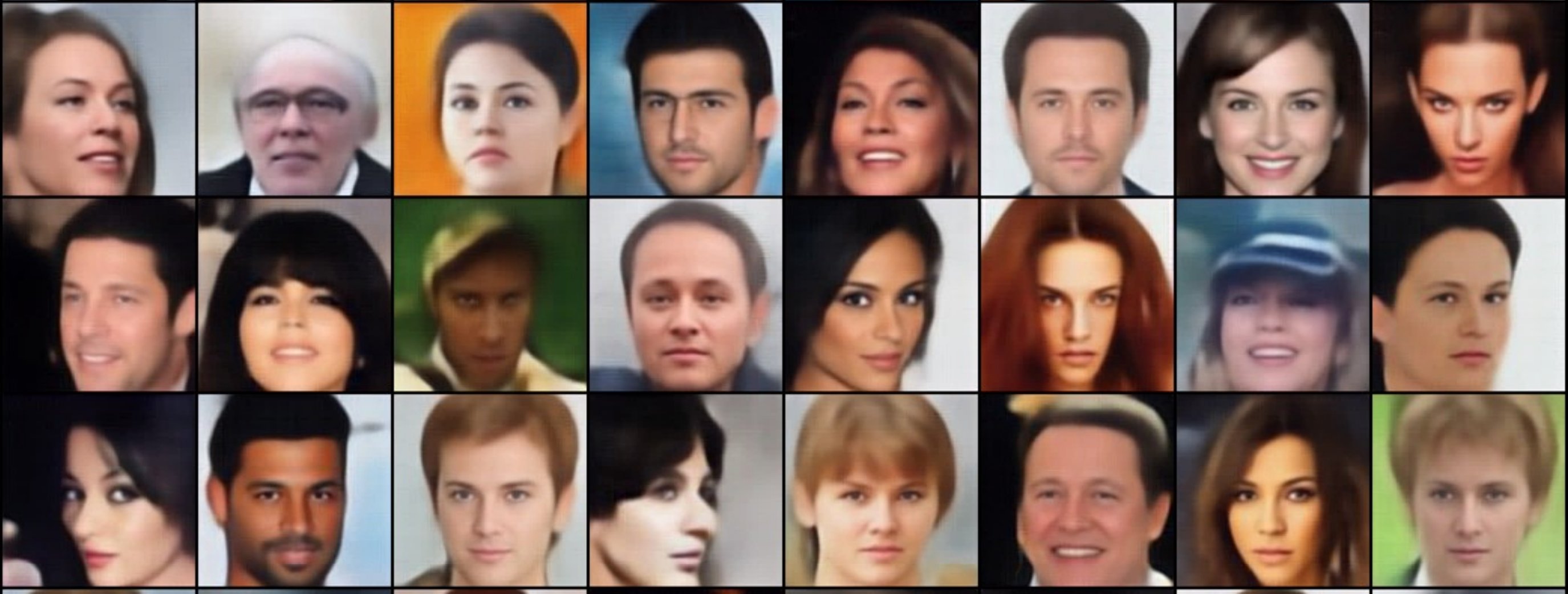}
    \caption[Generator result]{Images of fake faces generated using the generator of styleGAN-v2.}
    \label{fig:generator_result}
\end{figure}

\begin{figure}
    \centering
    \includegraphics[width=0.48\textwidth,trim=0.2cm 4cm 1cm 4cm, clip]{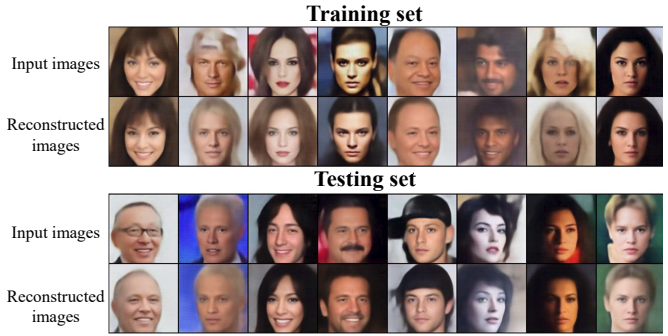}
    \caption[Generator result]{Images reconstructed from latent codes of CelebA images using generator of styleGAN-v2. Latent codes were computed using the facial image encoder. }
    \label{fig:encoder_and_generator_result}
\end{figure}

After training styleGAN-v2 with all CelebA data in 130,000 iterations with a batch size of 16, the resulting generator can generate clear fake facial images from random values, as shown in Fig.~\ref{fig:generator_result}. After training the facial image encoder, or GAN inverter, for 320,000 iterations with a batch size of 16, latent codes can be reconstructed into facial images similar to original facial images using the styleGAN-v2 generator with both training and testing sets (Fig.~\ref{fig:encoder_and_generator_result}).  

\subsection{Performance of Attribute Classifiers}

For each of the three binary attributes, an image classifier with ResNet-34 architecture was trained using all 29,264 images in CelebA to predict the attributes of facial images. The trained classifiers achieved 99.49\%\textendash 99.98\% and 80\%\textendash 95\% classification accuracy when classifying facial images belonging to training and testing sets, respectively (Table~\ref{tab:img_clsfer}). The classifier of male attributes achieved the highest accuracy in both training and testing sets. 

For each attribute, a linear SVM classifier was trained using latent codes computed from 70\% of CelebA images, in which 80\% of data were used for training and 20\% of data were to validate performance. Note that numbers of positive and negative samples were equal in training and validation sets. The accuracy of SVM classifiers in classifying attributes of validation data achieved 97.43\%, 96.58\%, and 87.54\% for male, wearing lipstick, and mouth slightly open, respectively (Table~\ref{tab:svm_rlt}). Performance of the three SVM classifiers decreased to 57.89\%\textendash 74.61\% when applied to latent codes of the remaining images.

For each attribute, fMRI attribute classifiers were trained using fMRI data corresponding to facial stimuli in the training set. The classifiers corresponding to "male", "mouth slightly open", and "wearing lipstick" achieved 74.62\%\textendash 78.60\% accuracy in the training set (Table~\ref{tab:fmri_clsfer_result}). In the testing set, classification accuracy was between 49.85\%\textendash 60.54\% and vote accuracy exceeded 65.00\%\textendash 75.00\%. Thus, deciding the attribute based on voting results was more stable and accurate than based on the classification result from a trial of fMRI data. 

\begin{table}
    \centering
    \caption[Boundary result]{Classification accuracy of SVM classifiers to predict attributes based on latent codes}
    \begin{tabular}{|l|c|c|}
    \hline
    \multirow{2}{*}{\textbf{Attribute}}  & {\textbf{Validation data}} & {\textbf{Testing data}} \\ 
     \cline{2-3}
                                  & {\textbf{\textit{Acc (\%)}}} & {\textbf{\textit{Acc (\%)}}}   \\ \hline
    Male                & 97.43                               & 74.61                              \\ \hline
    Mouth slightly open & 87.54                               & 57.89                              \\ \hline
    Wearing lipstick    & 96.58                               & 66.36                              \\ \hline
    \end{tabular}
    \label{tab:svm_rlt}
\end{table}

\begin{table}
    \centering
    \caption[fMRI attribute classifier result]{Classification accuracy of fMRI-based attribute classifiers}
    \begin{tabular}{|l|c|c|c|}
    \hline
    \multirow{2}{*}{\textbf{Attribute}} & \textbf{Training data} & \multicolumn{2}{c|}{\textbf{Testing data}} \\ 
    \cline{2-4}
      & \textbf{\textit{Acc (\%)}} & \textbf{\textit{Acc (\%)}} & \textbf{\textit{Vote acc (\%)}} \\ \hline
    Male                 & 74.62 & 52.87  & 70.00  \\ \hline
    Mouth slightly open  & 75.67 & 49.83  & 65.00  \\ \hline
    Wearing lipstick     & 78.60  & 60.54  & 75.00 \\ \hline
    \end{tabular}
    \label{tab:fmri_clsfer_result}
\end{table}

\subsection{Manipulating Attributes of Facial Images}

\begin{figure*}[t]
    \centering
    \includegraphics[width=0.8\textwidth,trim=0cm 4.85cm 0cm 7.5cm, clip]{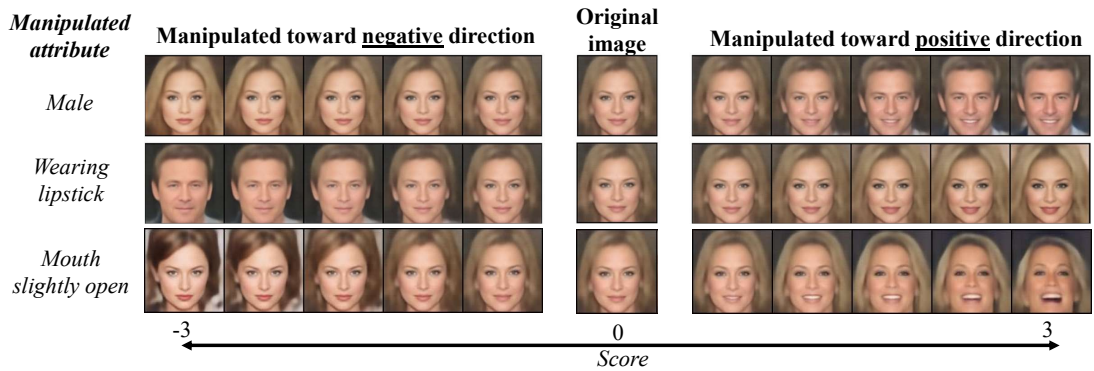}
    \caption[Manipulation of image attribute]{Manipulation of one attribute toward positive and negative directions by changing the distance between latent codes of images and boundary of the targeted attribute. For each attribute, 10 images were reconstructed from latent codes manipulated with scores ranging from -3 to 3. The latent codes were obtained from CelebA images using an encoder.}
    \label{fig:manipulated_images}
\end{figure*}

Using scores varying from -3 to 3 to manipulate latent codes, images generated using the StyleGAN-v2 generator can produce facial images with more negative or positive attributes relative to the original images (Fig.~\ref{fig:manipulated_images}). When manipulating attributes of ``male", the facial images corresponding to negative scores showed characteristics of female, whereas those corresponding to positive scores showed characteristics of male. However, when manipulating attributes of ``wearing lipstick", the negative scores produced facial images with enhanced attributes in both ``male" and ``no lipstick".   

\subsection{Reconstruction with and without Attribute Manipulation}

\begin{figure}
    \centering
    \includegraphics[width=0.48\textwidth,trim=4cm 3.4cm 2.5cm 3cm, clip]{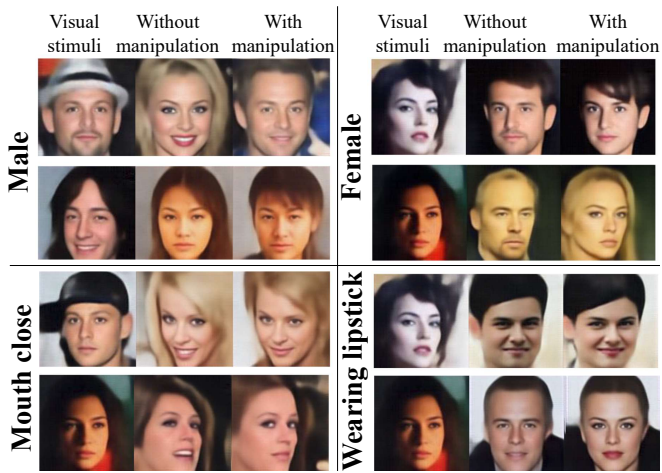}
    \caption[Manipulation of attributes of reconstructed images]{Images reconstructed from brain data in response to visual stimuli with and without attribute manipulation. Vvisual stimuli were with attributes of "male", "female", ``mouth close", or "wearing lipstick". Without manipulation, reconstructed images show incorrect attributes, but the manipulation mechanism successfully corrects the attributes.}
    \label{fig:Manipulate_recon_images}
\end{figure}

When applying image classifiers to predict attributes of images reconstructed from original mapped latent codes, classification accuracy was between 49.15\%\textendash 57.49\% (Table~\ref{tab:recon_attr_acc}). With attribute manipulation. Classification accuracy of images reconstructed from manipulated latent codes increased to 57.38\%\textendash65.27\%. The attribute manipulation mechanism appeared to enhance or correct attributes of reconstructed images (Fig.~\ref{fig:Manipulate_recon_images}), with accuracy improved between 3\%\textendash 8\%.         

\begin{table}[t]
    \centering
    \caption[Reconstruction attribute accuracy]{Attribute classification accuracy for images reconstructed from latent codes with and without attribute manipulation}
    \begin{tabular}{|l|c|c|}
    \hline
 \multirow{2}{*}{\textbf{Attribute}}  & {\textbf{No manipulation}} & {\textbf{With manipulation}} \\ 
     \cline{2-3}
                         &  {\textbf{\textit{Acc (\%)}}} &  {\textbf{\textit{Acc (\%)}}}   \\ \hline
                                  
    Male                 & 49.15 & 57.38  \\ \hline
    Mouth slightly open  & 54.79 & 57.83  \\ \hline
    Wearing lipstick     & 57.49 & 65.27 \\ \hline
    \end{tabular}
    \label{tab:recon_attr_acc}
\end{table}

\subsection{Reconstruction using the Proposed Method and VAE-GAN}
\begin{figure}
    \centering
    \includegraphics[width=0.48\textwidth,trim=2.2cm 6.8cm 2.5cm 6.2cm, clip]{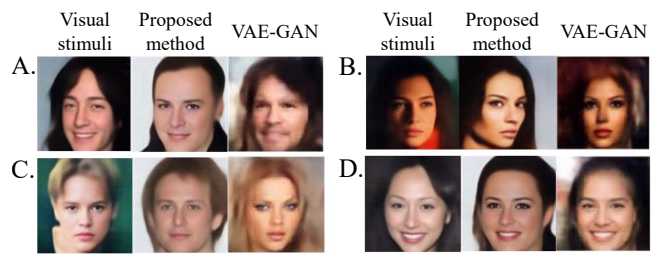}
    \caption[Comparison with VAE-GAN]{Comparison of Images reconstructed from fMRI brain activity using the proposed method and VAE-GAN \cite{VanRullen2019}.}
    \label{fig:method_compare}
\end{figure}

Compared with images reconstructed using VAE-GAN, images reconstructed using the proposed method were sharper (Fig.~\ref{fig:method_compare}). Moreover, in Fig.~\ref{fig:method_compare}C, the image reconstructed using VAE-GAN showed a female face, but the image reconstructed using the proposed method showed a male face, which is consistent with the attribute of the origin visual stimulus. 

\subsection{Evaluation of Reconstructed Images based on Human Judgement}
The variance for all questionnaire feedback is 0.00714, implying high consistency between different responses. Results of attribute decision showed the highest scores for ``male" attribute, 74.60\% of average accuracy and 75\% of vote accuracy (Table~\ref{tab:rec_attr_consistency_human}). Moreover, feedback variance for the male attribute (0.049) was the lowest among the three attributes, implying that the characteristic of gender was efficiently reconstructed in the images. In the 2AFC test, feedback showed 76.19\% average accuracy and 90.00\% vote accuracy. The feedback variance of the 2AFC test (0.159) was higher than that for attribute decision. 

\begin{table}
    \centering
    \caption[Human judgment]{Human judgment results for attributes of reconstructed images.}
    \begin{tabular}{|l|c|c|c|}
    \hline
    \multirow{2}{*}{{\textbf{Attribute}}}  &  \multicolumn{2}{c|}{\textbf{Performance}} & {\textbf{Feedback}}  \\ 
                         \cline{2-3}
                         & {\textbf{\textit{Average Acc (\%)}}} & {\textbf{\textit{Vote Acc (\%)}}}  & {\textbf{variance}} \\ \hline
    Male                 & 74.60            & 75.00         & 0.049 \\ \hline
    Mouth slightly open  & 67.57            & 71.66         & 0.055 \\ \hline
    Wearing lipstick     & 68.53            & 71.66         & 0.081 \\ \hhline{|=|=|=|=|}
    Average     & 70.23            & 72.77         & 0.062 \\ \hline
    
    \end{tabular}
    \label{tab:rec_attr_consistency_human}
\end{table}

\section{Discussion}\label{sec:discussion}
Reconstruction of visual stimuli from brain activity is a kind of mind reading. 
Taking advantage of the semantics of latent codes refers to knowledge acquired using in-domain GAN \cite{zhu2020domain}, the proposed framework can successfully reconstruct facial images with correct facial attributes. 
Moreover, utilizing the power of GAN, the proposed method produced sharper facial images.      
A comprehensive examination of the reconstruction performance revealed the high consistency between the attributes of reconstructed and original images and high differentiation between reconstructed images corresponding to different stimuli. These results support our idea that constraining latent space with semantics and conducting semantic editing enhances the quality of reconstructed images in regard to attributes, which leads to  better mind reading. 

In this study, we convert a reconstruction problem into a GAN inversion task.
This work seeks to relieve the bottleneck and enhance performance using in-domain GAN \cite{zhu2020domain}, which not only reconstructs facial images from brain activity, but also reveals semantics of inverted code.
We found that attribute manipulation can be used to improve consistency between the reconstructed image and the stimulus image (Tables~\ref{tab:recon_attr_acc}).
For example, comparing with VAE-GAN \cite{VanRullen2019}, we can retain the correct gender characteristic (Fig.~\ref{fig:method_compare}C) using the proposed framework.
Also, Table \ref{tab:rec_attr_consistency_human} shows that images generated by our method coincide with human perception of specific attributes.
These findings suggest that we can achieve better results by leveraging capabilities of learning semantically meaningful latent codes and adjusting facial characteristics.  


The proposed framework divides the problem of visual reconstruction into multiple sub-problems, each of which can be solved using current state-of-the-art methods.
Using styleGAN-v2 and in-domain GAN, the trained facial image encoder and generator reconstructed CelebA faces with high similarity to the original images (Fig.~\ref{fig:encoder_and_generator_result}). 
However, the similarity between the visual stimuli and images reconstructed from fMRI data became relatively low (Fig.~\ref{fig:Manipulate_recon_images}).
This may have resulted from the linear mapping between fMRI data and the latent space, which is a compromise solution to link brain activity and image latent space. 
Linear mapping does not consider knowledge of semantics. 
However, if linear mapping is replaced with deep learning approaches, insufficient fMRI data leads to over-fitting. 
Although fMRI-based and latent-code-based attribute classifiers (Tables~\ref{tab:fmri_clsfer_result} and \ref{tab:svm_rlt}) both showed much lower testing accuracy than image-based attribute classifiers (Table~\ref{tab:img_clsfer}), improving their accuracy may not increase the similarity between reconstructed and original facial images.
This is because manipulation of attributes generally modifies details of facial images and rarely changes the contour or outline of faces (Figs.~\ref{fig:manipulated_images} and \ref{fig:Manipulate_recon_images}).  
On the other hand, convolutional neural networks and SVM are powerful methods for classification and boundary decision, respectively. 
The low performance of the fMRI attribute classifier may result from insufficient data. 
Compared to the number of CelebA images for training image attribute classifiers, far less fMRI data was used to train attribute classifiers, which may have limited its performance. 
Thus, these results suggest that increasing the amount of fMRI data and replacing linear mapping with other advanced methods may further improve the performance of visual reconstruction.

In this study, activity in the occipital cortex and part of the temporal and parietal regions was selected for reconstruction of images. 
Compared to using only temporal or frontoparietal regions, reconstruction using brain activity in occipital regions achieves the highest performance \cite{VanRullen2019}. 
The occipital areas are mainly to process low-level features. 
To enrich information related to attributes, we additionally selected brain areas higher in the visual processing hierarchy, which is responsible for processing abstract concepts or categorical features.
The fusiform region, which is located in the inferior temporal cortex, serves an important function in face recognition \cite{mccarthy1997face}. 
Moreover, the superior parietal lobule and postcentral gyrus are on the visual ``where" pathway and are responsible for spatial attention \cite{caminiti1996sources,Iwamura2001}.        

Even though our comprehensive analyses show improved attribute consistency, this study has several limitations.  
First, the training data require annotations of attributes. 
Fortunately, CelebA provides 40 annotations for facial images. 
To label attributes of new datasets, semi-supervised learning largely reduces manual labeling efforts.
Second, manipulation of attributes may be biased by training data; thus, reconstructed images deviate from originals.
The second row of Fig.~\ref{fig:manipulated_images} shows that the manipulation in the positive direction of ``wearing lipstick" also enhanced female attributes.
However, when manipulating the image toward negative attributes such as ``female", ``not wearing lipstick", and ``mouth closed", the results might be biased because the original image was consistent with these attributes.
Fig.~\ref{fig:Manipulate_recon_images} also shows that a female with long hair became a male with short hair after manipulation, even though the original stimulus was a male with long hair. 
It is resulted from the co-occurrence of ``female" and ``wearing lipstick" and of ``male" and ``short hair" in the training data. 
Thus, although the attribute manipulation following \cite{shen2020interpreting} applied conditional manipulation to isolate one attribute from another, some entangled attributes are difficult to manipulate independently.
Third, images reconstructed from fMRI data corresponding to the same visual stimulus were different (Figs.~\ref{fig:Manipulate_recon_images} and \ref{fig:method_compare}).
To the best of our knowledge, previous visual reconstruction studies have not attempted to solve this many-to-one problem. 
However, we speculate that as the number of attributes to be manipulated increases and the performance of the fMRI-based attribute classifier is improved, the similarity among reconstructed images corresponding to a single stimulus will also increase.
Despite the limitations of this study, our results indicate that the proposed framework successfully improved attribute consistency.    

To conclude, this paper presents the first framework utilizing the power of GAN-inversion to edit attributes for fMRI-based visual reconstruction research. 
Our results further demonstrate that the proposed framework can reconstruct images from fMRI data having attributes highly consistent with the original visual stimuli.
We also indicated potential directions to improve the performance of the proposed framework in the future. 
Altogether, the proposed framework may shed light on visual reconstruction from brain activity in regard to attribute consistency.

\section*{Acknowledgment}
This work was supported in part by Ministry of Science and Technology, Taiwan (MOST-108-2221-E009-066-MY3 and MOST-109-2321-B-009-007) and Ministry of Education, Culture, Sports, Science and Technology, Japan (No. 20299115).
We are grateful to the National Center for High-performance Computing, Taiwan, for providing computing services and facilities.


\bibliographystyle{IEEEtran}
\bibliography{IEEEabrv,V1}

\begin{thebibliography}{10}
\providecommand{\url}[1]{#1}
\csname url@samestyle\endcsname
\providecommand{\newblock}{\relax}
\providecommand{\bibinfo}[2]{#2}
\providecommand{\BIBentrySTDinterwordspacing}{\spaceskip=0pt\relax}
\providecommand{\BIBentryALTinterwordstretchfactor}{4}
\providecommand{\BIBentryALTinterwordspacing}{\spaceskip=\fontdimen2\font plus
\BIBentryALTinterwordstretchfactor\fontdimen3\font minus
  \fontdimen4\font\relax}
\providecommand{\BIBforeignlanguage}[2]{{%
\expandafter\ifx\csname l@#1\endcsname\relax
\typeout{** WARNING: IEEEtran.bst: No hyphenation pattern has been}%
\typeout{** loaded for the language `#1'. Using the pattern for}%
\typeout{** the default language instead.}%
\else
\language=\csname l@#1\endcsname
\fi
#2}}
\providecommand{\BIBdecl}{\relax}
\BIBdecl

\bibitem{Kay2008}
K.~N. Kay, T.~Naselaris, R.~J. Prenger, and J.~L. Gallant, ``{Identifying
  natural images from human brain activity},'' \emph{Nature}, vol. 452, no.
  7185, pp. 352--355, 2008.

\bibitem{mozafari2020reconstructing}
M.~Mozafari, L.~Reddy, and R.~VanRullen, ``Reconstructing natural scenes from
  fmri patterns using bigbigan,'' in \emph{2020 International joint conference
  on neural networks (IJCNN)}.\hskip 1em plus 0.5em minus 0.4em\relax IEEE,
  2020, pp. 1--8.

\bibitem{Reddy2010}
L.~Reddy, N.~Tsuchiya, and T.~Serre, ``{Reading the mind's eye: Decoding
  category information during mental imagery},'' \emph{Neuro{I}mage}, vol.~50,
  no.~2, pp. 818--825, 2010.

\bibitem{takada2022generating}
S.~Takada, R.~Togo, T.~Ogawa, and M.~Haseyama, ``Generating captions of
  imagined content from human brain activities applying an image captioning
  model,'' in \emph{2022 IEEE 4th Global Conference on Life Sciences and
  Technologies (LifeTech)}.\hskip 1em plus 0.5em minus 0.4em\relax IEEE, 2022,
  pp. 614--615.

\bibitem{chang2021decoding}
P.-C. Chang, J.-R. Chang, P.-Y. Chen, L.-K. Cheng, J.-C. Hsieh, H.-Y. Yu, L.-F.
  Chen, and Y.-S. Chen, ``Decoding neural representations of rhythmic sounds
  from magnetoencephalography,'' in \emph{IEEE International Conference on
  Acoustics, Speech and Signal Processing (ICASSP)}.\hskip 1em plus 0.5em minus
  0.4em\relax IEEE, 2021, pp. 1280--1284.

\bibitem{Horikawa2013}
T.~Horikawa, M.~Tamaki, Y.~Miyawaki, and Y.~Kamitani, ``{Neural decoding of
  visual imagery during sleep},'' \emph{Science}, vol. 340, no. 6132, pp.
  639--642, 2013.

\bibitem{Owen2006}
A.~M. Owen, M.~R. Coleman, M.~Boly, M.~H. Davis, S.~Laureys, and J.~D. Pickard,
  ``{Detecting awareness in the vegetative state},'' \emph{Science}, vol. 313,
  no. 5792, p. 1402, 2006.

\bibitem{Schalk2004}
G.~Schalk, D.~McFarland, T.~Hinterberger, N.~Birbaumer, and J.~Wolpaw,
  ``{BCI}2000: A general-purpose brain-computer interface ({BCI}) system,''
  \emph{IEEE Transactions on Biomedical Engineering}, vol.~51, no.~6, pp.
  1034--1043, 2004.

\bibitem{fang2020reconstructing}
T.~Fang, Y.~Qi, and G.~Pan, ``Reconstructing perceptive images from brain
  activity by shape-semantic {GAN},'' \emph{Advances in Neural Information
  Processing Systems (NIPS)}, vol.~33, 2020.

\bibitem{lin2019dcnn}
Y.~Lin, J.~Li, and H.~Wang, ``{DCNN-GAN}: Reconstructing realistic image from
  f{MRI},'' in \emph{16th International Conference on Machine Vision
  Applications (MVA)}, 2019, pp. 1--6.

\bibitem{st2018generative}
G.~St-Yves and T.~Naselaris, ``Generative adversarial networks conditioned on
  brain activity reconstruct seen images,'' in \emph{2018 IEEE International
  Conference on Systems, Man, and Cybernetics (SMC)}.\hskip 1em plus 0.5em
  minus 0.4em\relax IEEE, 2018, pp. 1054--1061.

\bibitem{shen2019end}
G.~Shen, K.~Dwivedi, K.~Majima, T.~Horikawa, and Y.~Kamitani, ``End-to-end deep
  image reconstruction from human brain activity,'' \emph{Frontiers in
  Computational Neuroscience}, vol.~13, p.~21, 2019.

\bibitem{shen2019deep}
G.~Shen, T.~Horikawa, K.~Majima, and Y.~Kamitani, ``Deep image reconstruction
  from human brain activity,'' \emph{PLoS Computational Biology}, vol.~15,
  no.~1, p. e1006633, 2019.

\bibitem{horikawa2017generic}
T.~Horikawa and Y.~Kamitani, ``Generic decoding of seen and imagined objects
  using hierarchical visual features,'' \emph{Nature Communications}, vol.~8,
  no.~1, pp. 1--15, 2017.

\bibitem{VanRullen2019}
R.~VanRullen and L.~Reddy, ``Reconstructing faces from {fMRI} patterns using
  deep generative neural networks,'' \emph{Communications Biology}, no.~2, p.
  193, 2019.

\bibitem{donahue2019large}
J.~Donahue and K.~Simonyan, ``Large scale adversarial representation
  learning,'' \emph{arXiv preprint arXiv:1907.02544}, 2019.

\bibitem{liu2015faceattributes}
Z.~Liu, P.~Luo, X.~Wang, and X.~Tang, ``Deep learning face attributes in the
  wild,'' in \emph{Proceedings of International Conference on Computer Vision
  (ICCV)}, December 2015.

\bibitem{yan2016dpabi}
C.-G. Yan, X.-D. Wang, X.-N. Zuo, and Y.-F. Zang, ``{DPABI}: Data processing \&
  analysis for (resting-state) brain imaging,'' \emph{Neuroinformatics},
  vol.~14, no.~3, pp. 339--351, 2016.

\bibitem{Friston2007SPM}
K.~Friston, J.~Ashburner, S.~Kiebel, T.~Nichols, and W.~Penny, ``Statistical
  parametric mapping,'' in \emph{Academic Press}, London, 2007.

\bibitem{Ashburner2007dartel}
J.~Ashburner, ``A fast diffeomorphic image registration algorithm,''
  \emph{Neuro{I}mage}, vol.~38, no.~1, pp. 95--113, 2007.

\bibitem{Rolls2020aal}
E.~T. Rolls, C.-C. Huang, C.-P. Lin, J.~Feng, and M.~Joliot, ``Automated
  anatomical labelling atlas 3,'' \emph{Neuro{I}mage}, vol. 206, p. 116189,
  2020.

\bibitem{zhu2020domain}
J.~Zhu, Y.~Shen, D.~Zhao, and B.~Zhou, ``In-domain {GAN} inversion for real
  image editing,'' in \emph{European Conference on Computer Vision}.\hskip 1em
  plus 0.5em minus 0.4em\relax Springer, 2020, pp. 592--608.

\bibitem{shen2020interpreting}
Y.~Shen, J.~Gu, X.~Tang, and B.~Zhou, ``Interpreting the latent space of {GAN}s
  for semantic face editing,'' in \emph{Proceedings of the IEEE/CVF Conference
  on Computer Vision and Pattern Recognition (CVPR)}, 2020, pp. 9243--9252.

\bibitem{yang2021semantic}
C.~Yang, Y.~Shen, and B.~Zhou, ``Semantic hierarchy emerges in deep generative
  representations for scene synthesis,'' \emph{International Journal of
  Computer Vision}, vol. 129, no.~5, pp. 1451--1466, 2021.

\bibitem{karras2020analyzing}
T.~Karras, S.~Laine, M.~Aittala, J.~Hellsten, J.~Lehtinen, and T.~Aila,
  ``Analyzing and improving the image quality of {styleGAN},'' in
  \emph{Proceedings of the IEEE/CVF Conference on Computer Vision and Pattern
  Recognition (CVPR)}, 2020, pp. 8110--8119.

\bibitem{simonyan2014very}
K.~Simonyan and A.~Zisserman, ``Very deep convolutional networks for
  large-scale image recognition,'' \emph{arXiv preprint arXiv:1409.1556}, 2014.

\bibitem{Chang2011libsvm}
C.-C. Chang and C.-J. Lin, ``{LIBSVM}: A library for support vector machines,''
  \emph{ACM Trans. Intell. Syst. Technol.}, vol.~2, no.~3, may 2011.

\bibitem{mccarthy1997face}
G.~McCarthy, A.~Puce, J.~C. Gore, and T.~Allison, ``Face-specific processing in
  the human fusiform gyrus,'' \emph{Journal of Cognitive Neuroscience}, vol.~9,
  no.~5, pp. 605--610, 1997.

\bibitem{caminiti1996sources}
R.~Caminiti, S.~Ferraina, and P.~B. Johnson, ``The sources of visual
  information to the primate frontal lobe: a novel role for the superior
  parietal lobule,'' \emph{Cerebral Cortex}, vol.~6, no.~3, pp. 319--328, 1996.

\bibitem{Iwamura2001}
Y.~Iwamura, A.~Iriki, M.~Tanaka, M.~Taoka, and T.~Toda, ``{Processing of higher
  order somatosensory and visual information in the intraparietal region of the
  postcentral gyrus},'' in \emph{Somatosensory Processing From Single Neuron to
  Brain Imaging}, 2001.

\end{thebibliography}

\vspace{12pt}
\color{red}

\end{document}